\documentclass[aps, showpacs,pra,superscriptaddress,twocolumn,
eqsecnum] {revtex4}
\usepackage{hyperref}
\usepackage{amsmath }
\usepackage{amssymb}
\usepackage{epsfig}
\usepackage{natbib}
\newcommand*{\be}{\begin{equation}}
\newcommand*{\ee}{\end{equation}}
\newcommand*{\ud}{\mathrm{d}}
\begin{document}
\bibliographystyle{revtex}
\title{Perturbation theory for bright spinor
Bose--Einstein condensate solitons}

\author{Evgeny V. Doktorov}
\email{doktorov@dragon.bas-net.by}

\affiliation{B.I. Stepanov Institute of Physics, 220072 Minsk,
Belarus}

\author{Jiandong Wang}
\email{jwang@cems.uvm.edu}
\affiliation{Department of Mathematics
and Statistics, University of Vermont, Burlington, VT 05401}

\author{Jianke Yang}
\email{jyang@cems.uvm.edu}
\affiliation{Department of Mathematics
and Statistics, University of Vermont, Burlington, VT 05401}

\begin{abstract}
We develop a perturbation theory for bright solitons of the $F=1$
integrable spinor Bose-Einstein condensate (BEC) model. The
formalism is based on using the Riemann-Hilbert problem and
provides the means to analytically calculate evolution of the
soliton parameters. Both rank-one and rank-two soliton solutions
of the model are obtained. We prove equivalence of the rank-one
soliton and the ferromagnetic rank-two soliton. Taking into
account a splitting of a perturbed polar rank-two soliton into two
ferromagnetic solitons, it is sufficient to elaborate a
perturbation theory for the rank-one solitons only.
 Treating a small deviation from the integrability
condition as a perturbation, we describe the spinor BEC soliton
dynamics in the adiabatic approximation. It is shown that the
soliton is quite robust against such a perturbation and preserves
its velocity, amplitude, and population of different spin
components, only the soliton frequency acquires a small shift.
Results of numerical simulations agree well with the analytical
predictions, demonstrating only slight soliton profile
deformation.
\end{abstract}


\maketitle

\section{Introduction}

Bright and dark solitons in quasi-one-dimensional Bose--Einstein
condensates (BECs), observed experimentally
\cite{bright1,bright2,dark1,dark2}, are expected to be important
for various applications in atom optics \cite{book}, including
atom interferometry, atom lasers, and coherent atom transport.
Recent experimental and theoretical advances in BEC soliton
dynamics are reviewed in Refs. \cite{Brazh,Abdul,Carr}.

Spinor BEC of alkali atoms \cite{exp1,exp2} with a purely optical
confinement, along with the two-component condensate
\cite{Pan1,Valera, Hung}, represents an example of the condensate
with internal degrees of freedom which endow the solitons with
vectorial properties. Modulational instability in the spinor BEC
model was investigated in Ref. \cite{Robins}, and some exact
solutions and their stability were studied in Ref. \cite{Malomed}.
Vector gap solitons and self-trapped waves were identified in the
spinor BEC model loaded into one-dimensional optical lattice
potential \cite{Beata}. Recently bright-dark soliton complexes in
this model have been found \cite{Panos} by reducing it to the
completely integrable Yajima-Oikawa system \cite{YO}.

Wadati and co-workers found \cite{W1} that the three-component
nonlinear equations describing the BEC with the hyperfine spin
$F=1$ admit the reduction to another integrable model -- the
$2\times2$ matrix nonlinear Schr\"odinger (NLS) equation, after
imposing a constraint on the condensate parameters. Both bright
\cite{W2} and dark \cite{W3} solitons possessing properties of
true solitons of integrable equations have been found. The
formalism of the inverse scattering transform for the matrix NLS
equation under non-vanishing boundary conditions was developed in
Ref. \cite{W4} and extended in Ref. \cite{W5} to describe bright
spinor BEC soliton dynamics on a finite background. The full-time
description of the modulational instability development in the
integrable spinor BEC model was given both numerically
\cite{Malomed} and analytically \cite {DRK}.

Integrable models provide a very useful proving ground for testing
new analytical and numerical approaches to study such a
complicated system as the spinor BEC. At the same time, the
integrability conditions impose specific  restrictions on the
parameters of the model which can conflict with actual
experimental settings, despite the fact that the effective
interaction between atoms in BEC can be tuned, to some extent, by
the optically induced Feshbach resonance \cite{Fesh1,Fesh2}.
Besides, in experiment it is impossible to exactly hold the
conditions between parameters which assure integrability of the
model. Therefore, sufficiently general analytical results
concerning the full (nonintegrable) model with realistic
parameters would be of importance.

As a step in this direction, in the present paper we develop a
perturbation theory for the integrable spinor BEC model.
Evidently, small disturbance of the integrability condition can be
considered as a perturbation of the integrable model. Our
formalism is based on the Riemann--Hilbert (RH) problem associated
with the spinor BEC model. The main advantage of the proposed
method is its algebraic nature, as distinct from the method using
the Gel'fand--Levitan integral equations \cite{Tsuchida}. The
application of the RH problem for treating perturbed soliton
dynamics goes back to Refs. \cite{Yura1,me}. The modern version of
the perturbation theory in terms of the RH problem has been
developed in a series of papers \cite{Valera2,Vme1,Vme2,DMR,D},
with its most general formulation in Ref. \cite{Valera3}. Another
version of the soliton perturbation theory (the direct
perturbation theory)  has been developed on the basis of expanding
perturbed solutions into squared eigenfunctions of the linearized
soliton equations \cite{Kaup1990,ChenYang2002}.

As was shown by Wadati and co-workers \cite{W1,W2}, bright
solitons in the integrable spinor BEC model can exist in two spin
states -- ferromagnetic (non-zero total spin) and polar (zero
total spin). Energy of the polar soliton is greater than that of
the ferromagnetic soliton. Moreover, the polar soliton
demonstrates a two-humped profile in a wide range of its
parameters. Our numerical simulations revealed that the polar
soliton is unstable under the action of a perturbation of a rather
general form and splits into two ferromagnetic solitons. This fact
is crucial for the development of a perturbation theory for spinor
BEC solitons.

The paper is organized as follows. After formulating the model in
Sec. \ref{sec:model}, we introduce in Sec. \ref{sec:Jost} analytic
solutions of the associated spectral problem, in order to
formulate in Sec. \ref{sec:RH} the RH problem. Solving this
problem, we derive in Sec. \ref{sec:soliton} bright soliton
solutions of the integrable spinor BEC model, both for the
rank-one and rank-two projectors. The rank-one soliton is
characterized by the familiar hyperbolic secant profile, while the
rank-two soliton has a more complicated form \cite{W1}. Two types
of the rank-two solutions are exactly ferromagnetic and polar
solitons. We prove that the ferromagnetic rank-two soliton is
equivalent to the rank-one soliton. In virtue of the fact that the
perturbed polar soliton splits into two rank-two ferromagnetic
solitons, it is sufficient to develop a perturbation theory for
the rank-one soliton only. This is performed in Sec. \ref{sec:PT}.
We derive evolution equations for the soliton parameters which
exactly account for the perturbation and serve as the generating
equations for iterations. Section \ref{sec:adia} contains a
description of the soliton dynamics in the adiabatic approximation
of the perturbation theory. We show analytically that a
ferromagnetic soliton is quite robust against a small disturbance
of the integrability condition, the only manifestation of the
perturbation action is a minor shift of the soliton frequency.
Numerical simulations of the perturbed spinor BEC equations are in
close agreement with the analytical predictions revealing only a
small soliton shape distortion and little perturbation-induced
radiation. Section \ref{sec:conclusion} concludes the paper.

\section{Model}
\label{sec:model}

We consider an effective one-dimensional BEC trapped in a
pencil-shaped region elongated in the $x$ direction and tightly
confined in the transversal directions. The assembly of atoms in
the hyperfine spin $F=1$ state is described by a vector order
parameter ${\roarrow
\Phi}(x,t)=(\Phi_+(x,t),\Phi_0(x,t),\Phi_-(x,t))^T$, where its
components correspond to three values of the spin projection
$m_F=1,0,-1$. The functions $\Phi_\pm$ and $\Phi_0$ obey a system
of coupled Gross--Pitaevskii equations \cite{Suomi,W2}
\begin{eqnarray}\label{sys1}
i\hbar\partial_t\Phi_\pm=&-&\frac{\hbar^2}{2m}\partial_x^2\Phi_\pm+(c_0+c_2)(|\Phi_\pm|^2
+|\Phi_0|^2)\Phi_\pm\nonumber\\
&+&(c_0-c_2)|\Phi_\mp|^2\Phi_\pm+c_2\Phi_\mp^*\Phi_0^2,\\
i\hbar\partial_t\Phi_0=&-&\frac{\hbar^2}{2m}\partial_x^2\Phi_0+(c_0+c_2)(|\Phi_+|^2+|\Phi_-|^2)\Phi_0
\nonumber\\
&+&c_0|\Phi_0|^2\Phi_0+2c_2\Phi_+\Phi_-\Phi_0^*,\nonumber
\end{eqnarray}
where the constant parameters $c_0=(g_0+2g_2)/3$ and
$c_2=(g_2-g_0)/3$ control the spin-independent and spin-dependent
interaction, respectively. The coupling constant $g_f$ ($f=0,2$)
is given in terms of the $s$-wave scattering length $a_f$ in the
channel with the total hyperfine spin $f$,
\[
g_f=\frac{4\hbar^2a_f}{ma_\perp^2}\left(1-C\frac{a_f}{a_\perp}\right)^{-1}.
\]
Here $a_\perp$ is the size of the transverse ground state, $m$ is
the atom mass, and $C=-\zeta(1/2)\approx1.46$.

 It was noted in \cite{W1} that Eqs. (\ref{sys1}) are
 reduced to an integrable system under the constraint
 \be\label{const}
 c_0=c_2\equiv-c<0.
 \ee
 The negative $c_2$ means that we consider the ferromagnetic ground state of the
 spinor BEC with attractive interactions.
 The condition (\ref{const}), being written in terms of $g_f$ as
 $2g_0=-g_2>0$,
 imposes a constraint on the scattering lengths:
 $a_\perp=3Ca_0a_2/(2a_0+a_2)$. Redefining the function
 ${\roarrow \Phi}$ as
 ${\roarrow \Phi}\rightarrow(\phi_+,\sqrt{2}\phi_0,\phi_-)^T$,
 normalizing the coordinates as $t\rightarrow(c/\hbar)t$ and
 $x\rightarrow(\sqrt{2mc}/\hbar)x$, and accounting for the
 constraint (\ref{const}), we obtain a reduced system of equations
 in a dimensionless form:
 \be
 i\partial_t\phi_\pm+\partial_{x}^2\phi_\pm+2\left(|\phi_\pm|^2+2|\phi_0|^2\right)\phi_\pm+2\phi_
 \mp^*\phi_0^2=0\;,\label{sys2}
 \ee
 \[
 i\partial_t\phi_0+\partial_{x}^2\phi_0+2\left(|\phi_+|^2+|\phi_0|^2+|\phi_-|^2\right)\phi_0
 +2\phi_+\phi_0^*\phi_-=0\;.
 \]
 After arranging the components $\phi_\pm$ and $\phi_0$ into a
 $2\times2$ matrix $Q$,
 \be\label{Q}
 Q=\left(\begin{array}{cc}\phi_+ & \phi_0\\
 \phi_0& \phi_-\end{array}\right),
 \ee
 we transform
 Eqs. (\ref{sys2}) to the integrable matrix NLS
 equation
 \be
 i\partial_tQ+\partial_x^2Q+2QQ^\dag Q=0\;.\label{MNLS}
 \ee
 The matrix NLS equation (\ref{MNLS}) appears as
 a compatibility condition of the system of linear equations \cite{Tsuchida}
 \begin{eqnarray}
 \partial_x\psi&=&ik[\Lambda,\psi]+\hat Q\psi,\label{spectr}\\
 \partial_t\psi&=&2ik^2[\Lambda, \psi]+V\psi,\label{tempor}
 \end{eqnarray}
 where $\Lambda=\mathrm{diag}(-1,-1,1,1)$,
 \be
 \hat Q=\left(\begin{array}{cc}0 & Q\\
 -Q^\dag & 0\end{array}\right)\;,
 V=2k\hat Q+i\left(\begin{array}{cc} QQ^\dag & Q_x\\
 Q^\dag_x & -Q^\dag Q\end{array} \right)\;,\label{V}
 \ee
 and $k$ is a spectral parameter.
 Equation (\ref{spectr}) (the spectral problem) enables us to
 determine initial spectral data from the known potential $\hat Q_0$, while
 Eq. (\ref{tempor}) governs the temporal evolution of the spectral
 data. A new solution of Eq. (\ref{MNLS}) [and hence of the BEC
 equations (\ref{sys2})] is obtained as a result of the
 reconstruction of the potential $\hat Q$ from the time-dependent
 spectral data.

 \section{Jost and analytic solutions}
 \label{sec:Jost}

 To determine the spectral data, we introduce matrix Jost solutions $J_\pm(x,k)$ of the spectral
 problem (\ref{spectr}) by means of the asymptotes
 $J_\pm\to\openone$ as $x\to\pm\infty$. Since $\mathrm{tr}\Lambda=0$, we
 have $\det J_\pm=1$ for all $t$. Being solutions of the
 first-order equation (\ref{spectr}), the Jost functions are not
 independent but are interconnected by the scattering matrix $S$:
 \be\label{scatt}
 J_-=J_+ESE^{-1}, \qquad E=\exp(ik\Lambda x), \qquad \det S=1.
 \ee
 Besides, the Jost solutions and the scattering matrix obey the
 involution property. Indeed, since the potential $\hat Q$ is
 anti-Hermitian, we obtain
 \be\label{involJ}
 J_\pm^\dag(k^*)=J_\pm^{-1}(k).
 \ee
 Similarly for the scattering matrix:
 \be\label{involS}
 S^\dag(k)=S^{-1}(k).
 \ee
 Note that the scattering matrix is defined for real $k$.

 For the subsequent analysis, analytic properties of the Jost
 solutions are of primary importance. Let us represent the matrix
 Jost solution $J$ as a collection of columns:
 $J=(J^{[1]},J^{[2]},J^{[3]},J^{[4]})$, and consider the first column.
 Rewriting the spectral
 equation (\ref{spectr}) with the corresponding boundary
 conditions in the form of the Volterra integral equations, we
 obtain a closed system of equations for entries of the first
 column:
 \begin{eqnarray*}
 J_{-11}&=&1+\int_{-\infty}^x\ud
 x'(\phi_+J_{-31}+\phi_0J_{-41})(x'),\\
 J_{-21}&=&\int_{-\infty}^x\ud x'(\phi_0J_{-31}+\phi_-J_{-41})(x'),\\
 J_{-31}&=&-\int_{-\infty}^x\ud
 x'(\phi_+^*J_{-11}+\phi_0^*J_{-21})(x')e^{2ik(x-x')},\\
 J_{-41}&=&-\int_{-\infty}^x\ud
 x'(\phi_0^*J_{-11}+\phi_-^*J_{-21})(x')e^{2ik(x-x')}.
 \end{eqnarray*}
 The last two integrands point out that the column $J_-^{[1]}$ is
 analytic in the upper half-plane $\mathbb{C}_+$, where $\mathrm{Im}k>0$, and
 continuous on the real axis $\mathrm{Im}k=0$. This can be proved in
 the same way as for the scalar NLS equation, under the condition of
 sufficiently fast decrease of the potential $\hat Q$ at infinity.
 Similarly we
 obtain that the column $J_-^{[2]}$ is analytic in $\mathbb{C}_+$ as
 well, while the two other columns $J_-^{[3]}$ and $J_-^{[4]}$ are
 analytic in the lower half-plane $\mathbb{C}_-$ and continuous on
 the real axis $\mathrm{Im}k=0$. As regards the matrix solution
 $J_+$, its first and second columns $J_+^{[1]}$ and $J_+^{[2]}$
 are analytic in $\mathbb{C}_-$, while the third and forth ones
 $J_+^{[3]}$ and $J_+^{[4]}$ are analytic in $\mathbb{C}_+$.
 Therefore, the matrix function
 \be\label{psi+}
 \psi_+=\left(J_-^{[1]},J_-^{[2]},J_+^{[3]},J_+^{[4]}\right)
 \ee
 solves the spectral equation (\ref{spectr}) and is analytic as a
 whole in $\mathbb{C}_+$.

 It is not difficult to see from Eqs. (\ref{scatt}) and
 (\ref{psi+}) that the analytic solution $\psi_+$ can be expressed
 in terms of the Jost functions and some entries of the
 scattering matrix:
 \be\label{psi+Jost}
 \psi_+=J_+ES_+E^{-1}=J_-ES_-E^{-1},
 \ee
 where
 \be\label{mat}
 S_+(k)=\left(\begin{array}{cccc}s_{11} & s_{12} & 0 & 0\\ s_{21}
 & s_{22} & 0 & 0\\ s_{31} & s_{32} & 1 & 0\\ s_{41} & s_{42} & 0
 & 1\end{array}\right), \; S_-(k)=\left(\begin{array}{cccc} 1
 & 0 & s_{31}^* & s_{41}^*\\ 0 & 1 & s_{32}^* & s_{42}^* \\ 0 & 0
 & s_{33}^* & s_{43}^* \\ 0 & 0 & s_{34}^* & s_{44}^*
 \end{array}\right).
 \ee
 In writing the expression for $S_-$ we use the involution
 (\ref{involS}). These upper and lower block-triangular matrices $S_\pm$
 factorize the scattering matrix \cite{NMPZ}:
 $SS_-=S_+$. Besides, it follows from Eq. (\ref{psi+Jost}) and
 $\det J_\pm=1$ that
 \be\label{det+}
 \det\psi_+=m_+^{(2)}=m_-^{(2)*},
 \ee
 where $m_+^{(2)}$ ($m_-^{(2)}$) is the second-order principal upper (lower) minor
 of the scattering matrix.

 To obtain the analytic counterpart of $\psi_+$ in $\mathbb{C}_-$, we
 consider the adjoint spectral equation
 \be\label{adj}
 \partial_xK_\pm=ik[\Lambda,K_\pm]-K_\pm\hat Q
 \ee
 with the asymptotic conditions $K_\pm\to\openone$ at
 $x\to\pm\infty$. The inverse matrix $J^{-1}$ can serve as a
 solution of the adjoint equation (\ref{adj}). Now we write
 a closed system of integral equations for rows of the matrices
 $K_\pm$. For example, the first row $K_{-[1]}$ obeys the
 equations
 \begin{eqnarray*}
 K_{-11}&=&1+\int_{-\infty}^x\ud
 x'(\phi_+^*K_{-13}+\phi_0^*K_{-14})(x'),\\
 K_{-12}&=&\int_{-\infty}^x\ud
 x'(\phi_0^*K_{-13}+\phi_-^*K_{-14})(x'),\\
 K_{-13}&=&-\int_{-\infty}^x\ud
 x'(\phi_+K_{-11}+\phi_0K_{-12})(x')e^{-2ik(x-x')},\\
 K_{-14}&=&-\int_{-\infty}^x\ud
 x'(\phi_0K_{-11}+\phi_-K_{-12})(x')e^{-2ik(x-x')}.
 \end{eqnarray*}
 It is seen that the row
 $K_{-[1]}$ is analytic in $\mathbb{C}_-$. Similarly, the second
 row $K_{-[2]}$ is analytic in $\mathbb{C}_-$, too, and the rows
 $K_{-[3]}$ and $K_{-[4]}$ are analytic in $\mathbb{C}_+$. For the
 matrix solution $K_+$ we find that the rows $K_{+[1]}$ and
 $K_{+[2]}$ are analytic in $\mathbb{C}_+$, while $K_{+[3]}$ and
 $K_{+[4]}$ are analytic in $\mathbb{C}_-$. Therefore, the matrix
 function
 \be\label{psi-}
 \psi_-^{-1}=\left(K_{-[1]},K_{-[2]},K_{+[3]},K_{+[4]}\right)^T
 \ee
 solves the adjoint equation (\ref{adj}) and is analytic as a
 whole in $\mathbb{C}_-$. Similar to $\psi_+$, the function
 $\psi_-^{-1}$ is expressed in terms of the Jost solutions and the
 scattering matrix:
 \be\label{psi-Jost}
 \psi_-^{-1}=ET_+E^{-1}J_+^{-1}=ET_-E^{-1}J_-^{-1},
 \ee
 where the matrices $T_\pm$,
 \[
 T_+\!=\!\left(\begin{array}{cccc}s_{11}^* & s_{21}^* & s_{31}^* &
 s_{41}^*\\ s_{12}^* & s_{22}^* & s_{32}^* & s_{42}^*\\ 0 & 0 & 1
 & 0\\ 0 & 0 & 0 & 1\end{array} \right), \;
 T_-\!=\!\left(\begin{array}{cccc} 1 & 0 & 0 & 0\\0 & 1 & 0 & 0\\
 s_{31} & s_{32} & s_{33} & s_{34}\\ s_{41} & s_{42} & s_{43} &
 s_{44}\end{array}\right),
 \]
 provide one more factorization of the
 scattering matrix: $T_-=T_+S$.
 As in Eq. (\ref{det+}), we can write
 \be\label{det-}
 \det\psi_-^{-1}=m_+^{(2)*}=m_-^{(2)}.
 \ee
 Note that the analytic solutions
 satisfy the involution property as well:
 \be\label{invol}
 \psi_+^\dag(k)=\psi_-^{-1}(k^*).
 \ee
 This property can be taken as a definition of the analytic function
 $\psi_-^{-1}$ from the known analytic function $\psi_+$.

 \section{The Riemann-Hilbert problem}
 \label{sec:RH}

 Hence, we constructed two matrix functions $\psi_+$ and
 $\psi_-^{-1}$ which are analytic in complementary domains of the
 complex plane and conjugate on the real line. Indeed, it follows
 from Eqs. (\ref{psi+Jost}) and (\ref{psi-Jost}) that $\psi_\pm$
 obey the relation
 \be\label{RHP}
 \psi_-^{-1}(k)\psi_+(k)=EG(k)E^{-1}, \qquad \mathrm{Im}k=0,
 \ee
 where
 \be\label{G}
 G=T_+S_+=T_-S_-=\left(\begin{array}{cccc} 1 & 0 & s_{31}^* &
 s_{41}^*\\ 0 & 1 & s_{32}^* & s_{42}^*\\ s_{31} & s_{32} & 1 &
 0\\ s_{41} & s_{42} & 0 & 1\end{array}\right).
 \ee
 Equation (\ref{RHP}) determines a matrix Riemann-Hilbert problem,
 i.e. a problem of the analytic factorization of a nondegenerate
 matrix $G$ in (\ref{G}), given on the real line, into a product of two
 matrices which are analytic in complementary domains
 $\mathbb{C}_\pm$. The RH problem (\ref{RHP}) needs a normalization condition,
 which is usually taken as
 \be\label{normal}
 \psi_\pm(x,k)\to\openone \quad \mathrm{at} \quad |k|\to \infty.
 \ee

 The analytic matrix functions $\psi_\pm$ can be treated as a
 result of a nonlinear mapping between the potential $\hat Q(x)$
 and a set of the spectral data which uniquely characterizes a
 solution of the RH problem (\ref{RHP}) and (\ref{normal}). Conversely,
 the potential can be reconstructed from an asymptotic expansion
 of $\psi_\pm(x,k)$ for large $k$. Indeed, writing $\psi_\pm$ as
 \[
 \psi_+(x,k)=\openone+k^{-1}\psi_+^{(1)}+{\cal O}(k^{-2}),
 \]
 \[
 \psi_-^{-1}(x,k)=\openone+k^{-1}\psi_-^{(1)}+{\cal O}(k^{-2})
 \]
 and inserting these expansions into Eqs. (\ref{spectr}) and
 (\ref{adj}), we obtain
 \be\label{reconst}
 \hat Q=-i[\Lambda,\psi_+^{(1)}]=i[\Lambda,\psi_-^{(1)}].
 \ee
 Hence, having solved the RH problem, we can find solutions of the BEC
 equations.

 In general, the matrices $\psi_+$ and $\psi_-^{-1}$ can have
 zeros $k_j$ and $\kappa_l$ in the corresponding domains of
 analyticity: $\det\psi_+(k_j)=0$, $k_j\in\mathbb{C}_+$, and
 $\det\psi_-^{-1}(\kappa_l)=0$, $\kappa_l\in\mathbb{C}_-$. In
 virtue of the involution (\ref{invol}), we obtain $\kappa_l=k_l^*$
 and equal number $N$ of zeros in both half-planes. The
 corresponding RH problem is said to be nonregular, or the RH
 problem with zeros. They are zeros of the RH problem that
 determine soliton solutions of the BEC equations.
 It is seen from Eqs. (\ref{mat}) and (\ref{det+}) that
 zeros of $\psi_+$ nullify $2\times2$ minors of $\psi_+$. Hence, the
 rank of
 $\psi_+(k_j)$ can be equal to one or two. It means in turn that there exist one ($|1_j\rangle$)
 or two ($|1_j\rangle$ and $|2_j\rangle$)
 four-component eigenvectors
 that correspond to zero eigenvalue of $\psi_+(k_j)$:
 \begin{eqnarray}\label{aa}
 &\psi_+(k_j)|1_j\rangle=0 \qquad \mathrm{for \quad rank}\psi_+(k_j)=1,\\
 &\psi_+(k_j)|1_j\rangle=\psi_+(k_j)|2_j\rangle=0  \quad\mathrm{for \quad
 rank}\psi_+(k_j)=2.\nonumber
 \end{eqnarray}
 The geometric multiplicity of $k_j$ is equal to the
 dimension of the null space of $\psi_+(k_j)$ (1 or 2 in our case).
 In this paper, we only consider the case of zeros $k_j$ with its geometric
 multiplicity equal to the algebraic multiplicity [which is the order of the
 zero $k_j$ in $\det\psi_+(k)$]. Note that the solution of the RH problem
 for the general case of zeros with unequal geometric and algebraic
 multiplicities was elaborated in Ref. \cite{Val-J}.

 We will solve the matrix non-regular RH problem with zeros $k_1$ and $k_1^*$ by means of its
 regularization, i.e. by extracting from $\psi_+$ and $\psi_-^{-1}$ rational
 factors that are responsible for the appearance of zeros. Hence,
 $\det\psi_+(k_1)=0$ [and correspondingly
 $\det\psi_-^{-1}(k_1^*)=0$]. We need a rational matrix
 function $\Xi^{-1}(x,k)$ which has a pole in the point $k_1$.
 Let us take $\Xi^{-1}(x,k)$ in the form
 \[
 \Xi^{-1}(x,k)=\openone+\frac{k_1-k_1^*}{k-k_1}P^{(r)},
 \]
 where
 \be\label{proj}
 P^{(r)}=\sum_{l,m=1}^r|l\rangle(M^{-1})_{lm}\langle m|,
 \ee
 $\langle
 m|=|m\rangle^\dagger$ due to involution, and $r=\mathrm{rank}\,\psi_+(k_1)$.
 $P^{(r)}$ is a projector of rank $r$, $(P^{(r)})^2=P^{(r)}$,
 and entries of the $r\times r$ matrix $M$
 are determined by
 \[
 (M)_{lm}=\langle l|m\rangle=\sum_{a=1}^4(l)_a^*(m)_a.
 \]
 In the
 appropriate basis the projector is represented as $P^{(1)}=\mathrm{diag}(1,0,0,0)$
 or $P^{(2)}=\mathrm{diag}(1,1,0,0)$.
 This yields
 \[
 \det\Xi^{-1}=\left(\frac{k-k_1^*}{k-k_1}\right)^r.
 \]
 Therefore, the product $\psi_+(x,k)\Xi^{-1}(x,k)$ is regular in
 $k_1$. In the same way, the regularization of $\psi_-^{-1}$ in
 the point $k_1^*$ is performed by the rational function
 \be\label{xi}
 \Xi(x,k)=\openone-\frac{k_1-k_1^*}{k-k_1^*}P^{(r)},
 \ee
 which provides the product $\Xi\psi_-^{-1}$ to be regular in
 $k_1^*$. Therefore,
 the analytic functions are factorized as
 \be\label{reg}
 \psi_+(k)=\widetilde\psi_+(k)\Xi(k), \qquad
 \psi_-^{-1}(k)=\Xi^{-1}(k)\widetilde\psi_-^{-1},
 \ee
 with holomorphic functions $\widetilde\psi_\pm$ which determine the regular
 (without zeros) RH problem:
 \be\label{regRH}
 \widetilde\psi_-^{-1}(k)\widetilde\psi_+(k)=\Xi(k)EG(k)E^{-1}\Xi^{-1}(k),
 \quad k\in\mathrm{Re}.
 \ee

 For several pairs of zeros $(k_j,k_j^*)$, $j>1$,
 the regularization of the RH problem can be performed in the same
 step-by-step manner, with the appropriate definition of the eigenvectors within each
 step.
 However, for practical calculation
 of $N$-soliton effects it is much more convenient
 to expand a product of rational factors into simple fractions,
 thereby transforming the product-type expression into a sum-type
 one \cite{Vme2,Val-J}.

 It is easy to find the coordinate dependence of the eigenvectors.
 Indeed, differentiating (\ref{aa}) in $x$ and in $t$ with $j=1$ and accounting Eqs.
 (\ref{spectr}) and (\ref{tempor}) gives
 ($|l_1\rangle\equiv|l\rangle$)
 \be\label{xt}
 \partial_x|l\rangle=ik_1\Lambda|l\rangle,
 \quad \partial_t|l\rangle=2ik_1^2|l\rangle,
 \ee
 with $l=1$ for rank one, and $l=1,2$ for rank two.
 Hence,
 \be\label{tide}
 |l\rangle=\exp(ik_1\Lambda x+2ik_1^2\Lambda t)|
 l^{(0)}\rangle,
 \ee
 where $|l^{(0)}\rangle$ is the coordinate-free
 four-dimensional vector.

 Zeros $k_j$ and vectors $|l_j^{(0)}\rangle$ comprise the
 discrete data of the RH problem that determine the soliton
 content of a solution of the BEC equations. The continuous data
 are characterized by the off-block-diagonal parts of the matrix
 $G(k)$ (\ref{G}), $k\in\mathrm{Re}$, and are responsible for the
 radiation components. In the following section we concretize
 the above relations to obtain one-soliton solutions of Eqs.
 (\ref{sys2}).

 \section{Soliton solutions}
 \label{sec:soliton}
 \subsection{Rank-one soliton}

 To obtain the rank-one soliton solution of the BEC equations (\ref{sys2}),
 we consider the single pair $k_1$ and $k_1^*$ of zeros and
 the eigenvector $|1\rangle$. In accordance with
 Eq. (\ref{tide}), the eigenvector takes the form
 \begin{eqnarray}
 |1\rangle=\!\!&&(e^{-ik_1x-2ik_1^2t}n_1,\;e^{-ik_1x-2ik_1^2t}n_2,\nonumber\\
 &&e^{ik_1x+2ik_1^2t}n_3,\;
 e^{ik_1x+2ik_1^2t}n_4)^T,\label{1}
 \end{eqnarray}
 where $n_a$, $a=1,\ldots,4$ are complex numbers. The RH data are purely discrete: $N=1$,
 $G(k)=\openone$, $\widetilde\psi_\pm=\openone$. Hence, the solution of the
 RH problem is given by the rational function $\Xi$ in (\ref{xi})
 with the projector $P^{(1)}$. The reconstruction formula (\ref{reconst}) is simplified to
 \be\label{bb}
 \hat Q=-2\nu[\Lambda, P^{(1)}],
 \ee
 where we set $k_1=\mu+i\nu$ and the projector $P^{(1)}$
 in (\ref{proj}) is explicitly written as
 \begin{eqnarray*}
 P^{(1)}&=&\frac{1}{2}\left[\left(|n_1|^2+|n_2|^2\right)\left(|n_3|^2+|n_4|^2\right)
 \right]^{-1/2}\widetilde P\\
 &\times&e^{-2i\mu x-4i(\mu^2-\nu^2)t}\mathrm{sech}z.
 \end{eqnarray*}
 Here
 \[
 \widetilde P_{ab}=n_an_b^*, \; z=2\nu(x+4\mu t)+\rho, \;
 e^{2\rho}=\frac{|n_1|^2+|n_2|^2}{|n_3|^2+|n_4|^2}.
 \]
 Hence, it follows from Eq. (\ref{bb}) that the soliton solution is
 given by
 \be\label{solr1}
 Q=2\nu\Pi^{(1)}e^{-2i\mu x-4i(\mu^2-\nu^2)t}\mathrm{sech}z
 \ee
 with the polarization matrix
 \[
 \Pi^{(1)}\!\!=\frac{1}{2}\left[\left(|n_1|^2\!+|n_2|^2\right)\!\!\left(|n_3|^2\!+|n_4|^2\right)
 \right]^{-1/2}\!\!\left(\begin{array}{cc}n_1n_3^* & n_1n_4^* \\
 n_2n_3^* & n_2n_4^*\end{array}\right).
 \]
 Note that $n_2n_3^*=n_1n_4^*$ due to the structure of the matrix
 $Q$ in (\ref{Q}). Besides, the matrix $\Pi^{(1)}$ obeys automatically
 two conditions:
 \[
 \det\Pi^{(1)}=0, \quad
 |\Pi_{11}^{(1)}|^2+|\Pi_{22}^{(1)}|^2+2|\Pi_{12}^{(1)}|^2=1.
 \]

 Moreover, it is not difficult to show that the matrix $\Pi^{(1)}$
 depends only on two essential real parameters. Indeed, the rank-one
 soliton (\ref{solr1}) can be represented as
 \be\label{solr11}
 Q=2\nu\left(\begin{array}{cc}e^{-i\chi}\cos^2\theta &
 \cos\theta\sin\theta \\ \cos\theta\sin\theta &
 e^{i\chi}\sin^2\theta\end{array}\right)e^{i\varphi}\mathrm{sech}z,
 \ee
 where
 \[
 \cos\theta=\frac{|n_1|}{|n_1|^2+|n_2|^2}=\frac{|n_3|}{|n_3|^2+|n_4|^2},
 \quad \chi=\mathrm{arg}(n_3-n_4),
 \]
 \begin{eqnarray*}
 &&\varphi=-2\mu x-4(\mu^2-\nu^2)t+\phi_\alpha, \\
 &&\varphi_\alpha=\mathrm{arg}(n_1-n_4)=\mathrm{arg}(n_2-n_3).
 \end{eqnarray*}

 The soliton amplitude is determined by the parameter $\nu$, and its
 velocity is equal to $4\mu$. The parameters $\rho$ and
 $\phi_\alpha$ give the initial position of the soliton center and
 its initial phase, respectively. The angle $\theta$ determines the
 normalized population of atoms in different spin states, while the phase
 factor $e^{i\chi}$ is responsible for the relative phases between
 the components $\phi_\pm$ and $\phi_0$.

 It should be noted for future use that the constant soliton
 parameters acquire in general a slow $t$ dependence in the
 presence of perturbation. This results in a modification of the equations for coordinates:
 \begin{eqnarray}
 z&=&2\nu(x-\xi(t)), \quad \varphi=-\frac{\mu}{\nu}z+\delta(t),\\
 \xi(t)&=&-\frac{1}{2\nu}\left(8\int^t\ud
 t'\mu(t')\nu(t')+\rho(t)\right),\nonumber\\
 \delta(t)&=&-2\mu\xi(t)-4\int^t\ud
 t'[\mu^2(t')-\nu^2(t')]+\varphi_\alpha(t).\nonumber
 \end{eqnarray}

 \subsection{Rank-two soliton}

 As before, we begin with the pair $k_1$ and $k_1^*$ of zeros, but
 now we have two linearly-independent eigenvectors
 \begin{eqnarray}
 |1\rangle=\!\!&&(e^{-ik_1x-2ik_1^2t}p_1,\;e^{-ik_1x-2ik_1^2t}p_2,\nonumber\\
 &&e^{ik_1x+2ik_1^2t}p_3,\;
 e^{ik_1x+2ik_1^2t}p_4)^T,\label{12}\\
 |2\rangle=\!\!&&(e^{-ik_1x-2ik_1^2t}q_1,\;e^{-ik_1x-2ik_1^2t}q_2,\nonumber\\
 &&e^{ik_1x+2ik_1^2t}q_3,\;
 e^{ik_1x+2ik_1^2t}q_4)^T,\nonumber
 \end{eqnarray}
 with $p_a$ and $q_a$, $a=1,\ldots,4$, being
 complex numbers. The rational function $\Xi$ is given by Eq.
 (\ref{xi}) with the rank-two projector $P^{(2)}$. This projector
 is written in accordance with Eq. (\ref{proj}) as

 \begin{eqnarray*}
 P^{(2)}\!\!&=&\!\!\sum_{l,m=1}^2|m\rangle(M^{-1})_{ml}\langle l|
 =(\det
 M)^{-1}\\
 &\times&(M_{22}|1\rangle\langle1|-M_{12}|1\rangle\langle2|-M_{21}|2\rangle\langle1|+M_{11}
 |2\rangle\langle2|).
 \end{eqnarray*}
 In this case
 \[
 M=
 \left(\begin{array}{cc} A_1e^{z'}+B_1e^{-z'} & A_3e^{z'}+B_3e^{-z'} \\
 A_3^*e^{z'}+B_3^*e^{-z'} &
 A_2e^{z'}+B_2e^{-z'}\end{array}\right),
 \]
 $z'=2\nu(x+4\mu t)$, and
 \begin{eqnarray*}
 &&A_1=|p_1|^2+|p_2|^2, \quad B_1=|p_3|^2+|p_4|^2,\\
 &&A_2=|q_1|^2+|q_2|^2, \quad B_2=|q_3|^2+|q_4|^2,\\
 &&A_3=p_1^*q_1+p_2^*q_2, \quad B_3=p_3^*q_3+p_4^*q_4.
 \end{eqnarray*}
 Introducing the notations (to reproduce literally the results of Ref.
 \cite{W1})
 \begin{eqnarray}
 p_1q_2-p_2q_1\!\!&=&\!\!e^{\rho+i\sigma}, \; p_3q_2-p_2q_3=\beta^*,
 \;
 p_1q_4-p_4q_1=\gamma^*,\nonumber
 \\
 \label{not}
 &&p_1q_3-p_3q_1=p_4q_2-p_2q_4=\alpha^*,
 \end{eqnarray}
 we
 write explicitly the projector $P^{(2)}$ as
 \[
 P^{(2)}=\left(\begin{array}{cccc} P_{11} & P_{12} & P_{13} & P_{14}\\
 P_{12}^* & P_{22} & P_{14} & P_{24}\\ P_{13}^* & P_{14}^* &
 1-P_{11} & -P_{12}^*\\ P_{14}^* & P_{24}^* & -P_{12} &
 1-P_{22}\end{array}\right),\]
 where
 \begin{eqnarray}
 P_{11}&=&Z^{-1}(|\alpha|^2+|\gamma|^2+e^{2z}),\nonumber\\
 P_{22}&=&Z^{-1}(|\alpha|^2+|\beta|^2+e^{2z}),
 \nonumber\\
 P_{12}&=&-Z^{-1}(\alpha^*\beta+\alpha\gamma^*), \label{entries}\\
 P_{14}&=&e^{i\varphi}Z^{-1}(\alpha
 e^z-\alpha^*\mathcal{D}e^{-z}),\nonumber\\
 P_{13}&=&e^{i\varphi}Z^{-1}(\beta
 e^z+\gamma^*\mathcal{D}e^{-z}),\nonumber\\
 P_{24}&=&e^{i\varphi}Z^{-1}(\gamma
 e^z+\beta^*\mathcal{D}e^{-z}),\nonumber
 \end{eqnarray}
 \begin{eqnarray}
 &&\mathcal{D}=\det\Pi^{(2)},\quad\varphi=-2\mu
 x-4(\mu^2-\nu^2)t+\sigma,\nonumber\\
 \label{coor}
  &&\quad Z=\det M=1+e^{2z}+|\mathcal{D}|^2e^{-2z}.
 \end{eqnarray}
 $\Pi^{(2)}$ is the polarization matrix $\Pi^{(2)}=\left(\begin{array}{cc}\beta &
 \alpha\\ \alpha & \gamma\end{array}\right)$ subjected to the
 normalization condition \cite{W1}
 \be\label{norm}
 2|\alpha|^2+|\beta|^2+|\gamma|^2=1.
 \ee
 As a result, we immediately find from Eq. (\ref{bb}) with $P^{(2)}$ the
 rank-two soliton
 solution of the BEC equations (\ref{sys2}) \cite{W1}
 \be\label{sol}
 Q(x,t)=4\nu
 e^{i\varphi}Z^{-1}
 \left[\Pi^{(2)}
 e^z+\sigma_2\Pi^{(2)\dag}\sigma_2\mathcal{D}e^{-z}\right],
 \ee
 where $\sigma_2$ is the Pauli matrix.
 Notice that the soliton solution of the matrix NLS equation was
 previously obtained in Ref. \cite{Tsuchida} by means of the Gelfand--Levitan
 integral equations, while our derivation is purely algebraic.
 The soliton (\ref{sol}) was also derived
 by Gerdjikov and co-workers
 via the dressing procedure \cite{G}.

 We will distinguish between two featured cases of
 $\det\Pi^{(2)}$, namely, $\det\Pi^{(2)}=0$
 and $\det\Pi^{(2)}\ne0$. These cases display different spin properties.
 Indeed, the spin density vector
 ${\roarrow f}(x,t)=\mathrm{tr}(Q^\dag\vec{\sigma}Q)$, where
 ${\roarrow \sigma}$ is the set of the Pauli matrices, is given in
 general by a spatially odd function
 \begin{eqnarray}
 {\roarrow
 f}(x,t)=\left(\frac{4\nu}{Z}\right)^2\left(e^{2z}-|\mathcal{D}|^2e^{-2z}\right)\nonumber\\
 \times\left(\begin{array}{c}\alpha\bar\beta+\bar\alpha\beta+\alpha\bar\gamma+\bar\alpha\gamma\\
 i(\bar\alpha\beta-\alpha\bar\beta+\alpha\bar\gamma-\bar\alpha\gamma)\\
 |\beta|^2-|\gamma|^2\end{array}\right),\label{spin}
 \end{eqnarray}
 with absolute value being of the form
 \be\label{absval}
 |{\roarrow
 f}|=\left(\frac{4\nu}{Z}\right)^2\left|\,e^{2z}-|\mathcal{D}|^2e^{-2z}\right|\left(1
 -4|\mathcal{D}|^2\right)^{1/2}.
 \ee
 Therefore, the total spin vector $\roarrow F=\int\ud x{\roarrow f}(x,t)$ is zero.
 However, for $\mathcal{D}=0$, as it follows from Eq. (\ref{spin}),
 the absolute value of the total spin vector is nonzero,
 $|\roarrow F|=4\nu\ne0$.
 In accordance with this property, the case $\mathcal{D}=0$
 corresponds to the ferromagnetic state, while the case $\mathcal{D}\ne0$
 is usually referred to as a polar state. In fact, a true polar
 state corresponds to the condition $|\mathcal{D}|=1/2$, when, as it is seen from
 Eq. (\ref{absval}), the
 spin density is zero everywhere, not only the total spin \cite{W2}.

 It follows
 from Eq. (\ref{sol}) that the ferromagnetic state has the
 hyperbolic secant form
 \be\label{fer}
 Q^f=2\nu\,\Pi^{(2)} e^{i\varphi}\mathrm{sech}\,z,
 \ee
 where entries of the polarization matrix obey the normalization
 condition (\ref{norm}) and in addition the constraint
 $\beta\gamma-\alpha^2=0$. These two condition are sufficient to
 reduce the matrix $\Pi^{(2)}$ to the two-parameter form
 (\ref{solr11}) with the identifications
 \begin{eqnarray*}
 &&\cos\theta=\frac{|p_3|}{(|p_3|^2+|p_4|^2)^{1/2}}, \quad
 \chi=\mathrm{arg}(p_3-p_4),\\
 &&\varphi_\alpha=\mathrm{arg}(p_1-p_3)=\mathrm{arg}(p_4-p_2).
 \end{eqnarray*}
 Therefore, the rank-two ferromagnetic soliton is completely
 equivalent to the rank-one soliton (\ref{solr11}). Introducing the
 atom number density $n(x,t)$ and energy density $e(x,t)$,
 \[
 n(x,t)=\mathrm{tr}(Q^\dag Q), \quad e(x,t)=c\,\mathrm{tr}(Q_x^\dag
 Q_x-Q^\dag QQ^\dag Q),
 \]
 as well as their total counterparts $N_T=\int\ud xn(x,t)$ and
 $E_T=\int\ud xe(x,t)$, we obtain explicitly the total number of
 atoms and total energy in the ferromagnetic state:
 \[
 N_T^f=4\nu, \quad E_T^f=4cN_T^f(\mu^2-\nu^2/3).
 \]

 In turn, the total number of
 atoms in the polar state and its energy are given by
 \[
 N_T^p=8\nu, \qquad E_T^p=4cN_T^p(\mu^2-\nu^2/3).
 \]
 The energy difference between both states with equal amount of
 atoms is $E_T^f-E_T^p=-(1/16)c(N_T^f)^3<0$.
 Hence, the ferromagnetic state is energetically preferable, from the viewpoint
 of stability, as
 compared with the polar state.

 The atom number density of the polar soliton is described by the
 function
 \be\label{numden}
 n_p(z)=\left(\frac{4\nu}{Z}\right)^2\left(e^{2z}+4|\mathcal{D}|^2+|\mathcal{D}|^2e^{-2z}
 \right).
 \ee
 \begin{figure}
   \begin{center}
   \includegraphics[width=0.45\textwidth]{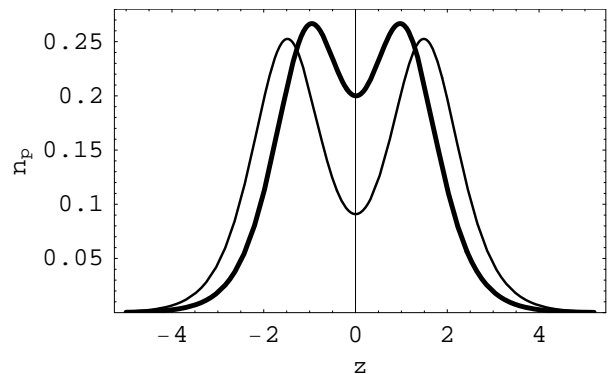}
   \caption{Profiles of the atom number density function of the polar
   state: $|\mathcal{D}|=1/8$ (thick line), $|\mathcal{D}|=1/20$ (thin
   line), $\nu=0.5$.}\label{fig1}
   \end{center}
   \end{figure}
 Figure \ref{fig1} demonstrates typical profiles of the atom
 number density function (\ref{numden}) for different
 $|\mathcal{D}|$. The two-humped structure becomes more pronounced
 with decreasing $|\mathcal{D}|$. Such a state can be treated as a
 pair of two ferromagnetic solitons with antiparallel spins
 \cite{W2}. Previous analysis of stability of multi-humped vector
 solitons for the cubic nonlinearity revealed that they are always
 unstable \cite{J1,J2}. Hence, we can suggest that the most likely
 scenario of the polar soliton evolution under the action of a
 perturbation would be its splitting into a pair of ferromagnetic
 solitons. Indeed, extensive simulations of the perturbed polar
 soliton behavior demonstrates unambiguously such a splitting. An
 example of such a behavior is depicted in Fig. \ref{fig2}, where
 we consider a disturbance of the integrability condition
 (\ref{const}) as a perturbation with a small parameter $\epsilon
 =c_0-c_2$ (see Eq. (\ref{pert}) below for a functional form of the
 perturbation). It is seen that all of the components of the polar
 soliton split under the action of the perturbation.
 \begin{figure}
   \begin{center}
   {\hspace{-2mm}\includegraphics[width=0.48\textwidth]{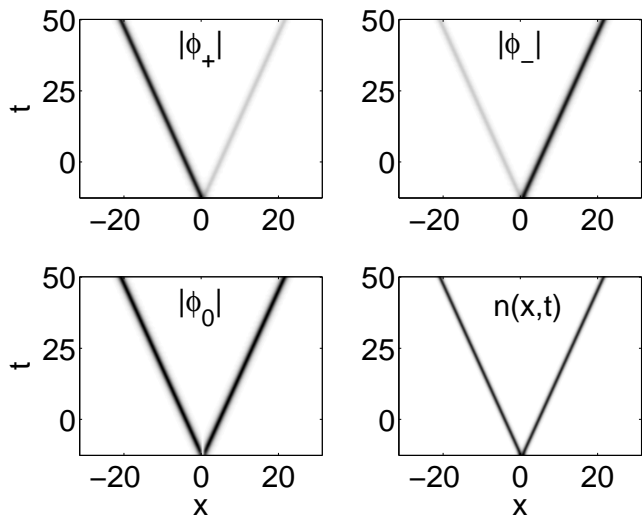}}
   \caption{Splitting of the components $\phi_\pm$ and $\phi_0$ of a perturbed polar soliton and
   of the atom number density $n(x,t)$. Here $|\beta|^2=0.7$,
   $|\alpha|^2=|\gamma|^2=0.1$. The perturbation is of the form
   (\ref{per-sys}) and (\ref{Rform}) with $\epsilon=0.1$.}\label{fig2}
   \end{center}
   \end{figure}

 Let us summarize the main conclusions concerning the soliton
 solutions which will play the key role in studying soliton
 perturbations. First, we derived the rank-one soliton solution
 with the hyperbolic secant profile. Second, rank-two solutions
 were obtained and classified as ferromagnetic and polar solitons.
 The polar soliton is perturbatively unstable and splits into two
 rank-two ferromagnetic solitons. Third, we proved equivalence of
 the rank-one soliton and rank-two ferromagnetic soliton.
 Therefore,
 it is sufficient to elaborate a perturbation
 theory for the more familiar type of solitons -- the rank-one
 soliton (\ref{solr11}). This will be done in the following section.

 \section{Perturbation theory for the bright spinor BEC soliton}
 \label{sec:PT}

 In this section we perform a general analysis of the perturbed
 spinor BEC equations
 \begin{eqnarray}
 i\partial_t\phi_\pm+\partial_{x}^2\phi_\pm&+&2\left(|\phi_\pm|^2
 +2|\phi_0|^2\right)\phi_\pm\nonumber\\
 &+&2\phi_
 \mp^*\phi_0^2=\epsilon R_\pm\;,\label{per-sys}
 \end{eqnarray}
 \begin{eqnarray*}
 i\partial_t\phi_0+\partial_{x}^2\phi_0&+&2\left(|\phi_+|^2+|\phi_0|^2+|\phi_-|^2\right)\phi_0
 \nonumber\\
 &+&2\phi_+\phi_0^*\phi_-=\epsilon R_0\;.\nonumber
 \end{eqnarray*}
 Here $R_\pm$ and $R_0$ determine a functional form of a
 perturbation, and $\epsilon $ is a small parameter. To
 distinguish between the `integrable' and `perturbative'
 contributions, we will assign the symbol
 $\delta/\delta t$ to the latter. Hence,
 \[
 i\frac{\delta\hat Q}{\delta t}=\epsilon\hat R, \quad \hat
 R=\left(\begin{array}{ll} 0 & R\\ R^\dag & 0\end{array}\right),
 \quad R=\left(\begin{array}{ll} R_+ & R_0\\ R_0 &
 R_-\end{array}\right).
 \]

 In general, a
 perturbation causes a slow evolution of the RH data. Indeed, a perturbation leads to a
 variation $\delta\hat Q$ of the potential entering the spectral
 equation (\ref{spectr}), and in turn to a variation of the Jost
 solutions:
 \[
 \delta J_{\pm x}=ik[\Lambda,\delta J_\pm]+\delta\hat QJ_\pm+\hat
 Q\delta J_\pm.
 \]
 Solving this equation  gives
 \[
 \delta J_\pm=J_\pm E\left(\int_{\pm\infty}^x\ud
 x'E^{-1}J_\pm^{-1}\delta\hat QJ_\pm E\right)E^{-1}.
 \]
 As a result, we find from Eqs. (\ref{scatt}), (\ref{psi+Jost}),
 and (\ref{psi-Jost}) a variation of the scattering matrix:
 \begin{eqnarray*}
 \frac{\delta S}{\delta t}&=&-i\epsilon S_+\int_{-\infty}^\infty\ud
 xE^{-1}\psi_+^{-1}\hat R\psi_+ES_-^{-1}\\
 &=&-i\epsilon T_+^{-1}\int_{-\infty}^\infty\ud
 xE^{-1}\psi_-^{-1}\hat R\psi_-ET_-.
 \end{eqnarray*}
 Here $S_\pm$ and $T_\pm$ are the matrices defined in Sect. \ref{sec:Jost}.
 Notice that they are the analytic solutions $\psi_\pm$ that enter
 naturally into this equation. Let us denote
 \begin{eqnarray}\label{upsilon}
 \Upsilon_\pm(a,b)&=&\int_a^b\ud xE^{-1}\psi_\pm^{-1}\hat R\psi_\pm
 E,\\
 \Upsilon_\pm(k)&\equiv&\Upsilon_\pm(-\infty,
 \infty).\nonumber
 \end{eqnarray}
 Then
 \[
 \frac{\delta S}{\delta t}=-i\epsilon
 S_+\Upsilon_+(k)S_-^{-1}=-i\epsilon T_+^{-1}\Upsilon_-(k)T_-.
 \]
 The matrices $\Upsilon_\pm$ are interrelated by means of the
 matrix $G$ entering the RH problem (\ref{RHP}):
 \be\label{up-up}
 \Upsilon_-(k)=G\Upsilon_+(k)G^{-1}.
 \ee
 Eventually, variations of the analytic solutions follow from Eqs.
 ( \ref{psi+Jost}) and (\ref{psi-Jost}):
 \[
 \frac{\delta\psi_+}{\delta t}=-i\epsilon\psi_+EH_+E^{-1}, \quad
 \frac{\delta\psi_-^{-1}}{\delta t}=i\epsilon
 EH_-E^{-1}\psi_-^{-1}.
 \]
 Here $H_\pm$ are the evolution functionals \cite{Valera2,Valera3} that
 are defined in terms of $\Upsilon_\pm$,
 \begin{eqnarray}\label{pi}
 H_+&=&\Upsilon_+(k)M_1-\Upsilon_+(x,\infty),\\
 H_-&=&M_1\Upsilon_-(k)-\Upsilon_-(x,\infty), \quad
 M_1=\mathrm{diag}(1,1,0,0),\nonumber
 \end{eqnarray}
 and contain all essential information about a perturbation. In
 particular, the evolution equations for $\psi_\pm$ gain
 additional terms caused by the perturbation and expressed in
 terms of $H_\pm$:
 \begin{eqnarray}\label{per-evol}
 \partial_t\psi_+&=&2ik^2[\Lambda,\psi_+]+V\psi_+-i\epsilon\psi_+EH_+
 E^{-1},\\
 \partial_t\psi_-^{-1}&=&2ik^2[\Lambda,\psi_-^{-1}]-\psi_-^{-1}V+i\epsilon
 EH_-E^{-1}\psi_-^{-1}.\nonumber
 \end{eqnarray}
 Besides, the evolution  equation for the matrix $G$ of the RH
 problem has the form
 \be\label{RH-cont}
 \partial_tG=2ik^2[\Lambda,G]-i\epsilon(GH_+-H_-G).
 \ee
 In fact, this equation gives the evolution of the continuous RH
 data.
 Note that the involution (\ref{invol}) connects $H_+$ with
 $H_-$:
 $
 H_-=H_+^\dag, \quad
 k\in\mathrm{Re}.
 $

 Now we consider a single rank-one soliton and derive perturbation-induced
 evolution equations for the discrete RH data,
 i.e. for the zero $k_1$ and the eigenvector $|1\rangle$.
 It is more convenient to work with the vector
 $|n\rangle=(n_1,n_2,n_3,n_4)^T$  which is
 constant in the absence of perturbation and
 acquires slow $t$ dependence under the action of a perturbation. We
 start from the equation
 \[
 \psi_+(k_1)|1\rangle=\psi_+(k_1)\exp\left[\left(ik_1x
 +2i\int\ud tk_1^2\right)\Lambda\right]|p\rangle=0
 \]
 which is valid irrespectively of the presence of a perturbation.
 Here the integral in the exponent accounts for a possible
 perturbation-induced time dependence of the zero $k_1$.
 Taking the total derivative in $t$, we obtain
 \begin{eqnarray*}
 &&\biggl(\partial_t\left[\psi_+(k)e^{ik\Lambda
 x+2i\int d t k_1^2\Lambda}\right]\\
 &&+\partial_k\left[\psi_+(k)e^{ik\Lambda
 x+2i\int d t k_1^2\Lambda}\right]\partial_tk\biggr)_{|k_1}\\
 &&+\psi_+(k_1)e^{ik_1\Lambda
 x+2i\int dt k_1^2\Lambda}\partial_t|p\rangle=0.
 \end{eqnarray*}
 The first term with $\partial_t\psi_+$ is given by Eq.
 (\ref{per-evol}) which contains the evolution functional $H_+$.
 Recall that
 the evolution functional $H_+(k)$ is defined in terms of
 $\Upsilon_+$ in (\ref{pi}) which in turn depends on $\psi_+^{-1}$.
 Hence, the function $H_+$ is meromorphic in $\mathbb{C}_+$
 with the simple pole in $k_1$, where $\psi_+$ has zero:
 \[
 H_+(k)=H_+^{(\mathrm{reg})}(k)+\frac{1}{k-k_1}\mathrm{Res}[H_+(k),k_1].
 \]
 Here $H_+^{(\mathrm{reg})}$ stands for the regular part of $H_+$ in the
 point $k_1$. Following now the method developed in Refs.
 \cite{Valera2,DMR,Valera3}, we find
 that the perturbed evolution of the vector $|n\rangle$ is given
 by
 \be\label{p-evol}
 \partial_t|n\rangle=i\epsilon e^{-2i\int\ud tk_1^2\Lambda}
 H_+^{(\mathrm{reg})}(k_1)e^{2i\int\ud tk_1^2\Lambda}|n\rangle.
 \ee
 Since the left-hand side of Eq. (\ref{p-evol}) is evidently
 $x$-independent, we can consider this equation for $x\to+\infty$,
 where $H_+$ has only two non-zero columns [see Eq. (\ref{pi})]:
 \be\label{H+}
 H_+(x\to+\infty)=\left(\Upsilon_+^{[1]},\Upsilon_+^{[2]},0,0\right).
 \ee
 Hence, Eqs. (\ref{p-evol}), written in components, take the form
 \begin{eqnarray}
 &\partial_tn_1=i\epsilon\left(X_{11}n_1
 +X_{12}n_2\right),\nonumber\\
 &\partial_tn_2=i\epsilon\left(X_{21}n_1
 +X_{22}n_2\right),\label{pp-evol}\\
 &\partial_tn_3=i\epsilon \left(X_{31}n_1
 +X_{32}n_2\right),\nonumber\\
 &\partial_tn_4=i\epsilon\left(X_{41}n_1
 +X_{42}n_2\right),\nonumber
 \end{eqnarray}
 where for simplicity we use the notation
 $X_{ab}=\Upsilon_{+ab}^{(\mathrm{reg})}(k_1)$, $a,b=1,2$,
 and $X_{ab}=\Upsilon_{+ab}^{(\mathrm{reg})}(k_1)e^{-4i\int\ud tk_1^2}$ for
 $a=3,4$ and $b=1,2$. Here
 $\Upsilon_+^{(\mathrm{reg})}$ is the
 regular part of $\Upsilon_+$ in the point $k_1$:
 \begin{eqnarray}\label{regul}
 \Upsilon_+^{(\mathrm{reg})}(k_1)&=&\int\ud xE^{-1}(k_1)\bigl\{\hat
 R(\openone-P^{(1)})\\
 &+&P^{(1)}\hat RP^{(1)}+2\nu x[\Lambda,P^{(1)}\hat
 R(\openone-P^{(1)})]\bigr\}E(k_1).\nonumber
 \end{eqnarray}
 This relation follows from $\Upsilon_+(k)=\int\ud
 xE^{-1}\Xi^{-1}\hat R\Xi E$.
 Note that in virtue of the
 specific symmetry of the matrix $\hat Q$ (\ref{V}) the entries
 $\Upsilon_{+32}$ and $\Upsilon_{+41}$ are equal, as well as the entries
 $\Upsilon_{+14}$ and $\Upsilon_{+23}$.

 Now we can derive the evolution equation for the parameters
 $\theta$ and $\chi$ entering the
 polarization
 matrix of the soliton solution (\ref{solr11}). Indeed, these
 parameters are defined in terms of $n_a$ which in turn obey Eqs.
 (\ref{pp-evol}). Simple calculation gives
 \begin{eqnarray}
 \partial_t\cos\theta&=&\frac{i\epsilon}{2}
 \biggl[e^{\rho+i\varphi_\alpha}\left(X_{31}e^{-i\chi}\cos\theta+X_{41}\sin\theta\right)\nonumber\\
 &-&e^{\rho-i\varphi_\alpha}\left(X_{31}^*e^{i\chi}\cos\theta+X_{41}^*\sin\theta\right)
 \biggr],\label{10-1}
 \end{eqnarray}
 \begin{eqnarray}
 &\partial_t&\!\!\!\chi=\frac{\epsilon}{2}\biggl[e^{\rho+i\varphi_\alpha}
 \bigl(X_{31}e^{-i\chi}-X_{42}e^{i\chi}
 +(\tan\theta-\cot\theta)
 X_{41}\bigr)\label{10-2}\nonumber\\
 &-&\!\!e^{\rho-i\varphi_\alpha}\bigl(X_{31}^*e^{i\chi}-X_{42}^*e^{-i\chi}
 +(\tan\theta-\cot\theta)
 X_{41}^*\bigr)\biggr].\label{10-2}
 \end{eqnarray}
 Just in the same way we obtain evolution equations for the
 parameters $\varphi_\alpha$ and $\rho$ which are also expressed in
 terms of $n_a$:
 \begin{eqnarray}
 \partial_t\varphi_\alpha&=&\frac{\epsilon}{2}\biggl[X_{11}+X_{11}^*+\left(X_{12}e^{-i\chi}+X_{12}^*
 e^{i\chi}\right)\tan\theta\nonumber\\
 &-&e^{\rho+i\varphi_\alpha}\left(X_{41}\cot\theta+X_{42}e^{i\chi}\right)\label{10-3}\\
 &-&e^{\rho-i\varphi_\alpha}\left(X_{41}^*\cot\theta+X_{42}^*e^{-i\chi}\right)\biggr],\nonumber
 \end{eqnarray}
 \begin{eqnarray}
 &\partial_t&\!\!\rho=\frac{i\epsilon}{2}\Bigl\{\left(X_{11}-X_{11}^*\right)\cos^2\theta
 +\left(X_{22}-X_{22}^*\right)\sin^2\theta\nonumber\\
 &+&\!\!\left[\left(X_{12}-X_{21}^*\right)e^{i\chi}-\left(X_{12}^*-X_{21}
 \right)e^{-i\chi}\right]\sin\theta\cos\theta\nonumber\\
 &+&e^{\rho+i\varphi_\alpha}(X_{31}e^{-i\chi}\cos^2\theta\label{10-4}\\
 &+&2X_{41}\sin\theta\cos\theta
 +X_{42}e^{i\chi}\sin^2\theta)\nonumber\\
 &-&e^{\rho-i\varphi_\alpha}(X_{31}^*e^{i\chi}\cos^2\theta\nonumber\\
 &+&2X_{41}^*\sin\theta\cos\theta
 +X_{42}^*e^{-i\chi}\sin^2\theta)\Bigr\}.\nonumber
 \end{eqnarray}
 In fact, Eqs. (\ref{10-1})--(\ref{10-4}) are greatly simplified when
 calculating the functions $X_{ab}$ for a specific perturbation.
 This will be demonstrated in the next Section.

 To derive evolution equation for $k_1$ (and hence for the soliton
 amplitude $\nu$ and velocity $\mu$), we start from the equation
 $\det\psi_+(k_1)=0$. Taking the total derivative in $t$ yields
 \[
 \partial_t\left(\det\psi_+(k)\right)_{|k_1}+\left(\partial_k\det\psi_+(k)\right)_{|k_1}
 \!\partial_tk_1=0.
 \]
 In accordance with Eqs. (\ref{reg}) and (\ref{xi}) we can write
 \[
 \det\psi_+(k)=\frac{k-k_1}{k-k_1^*}\det\tilde\psi_+(k),
 \]
 where $\det\tilde\psi_+(k_1)\ne0$ because $\tilde\psi_+(k)$ is a
 solution of the regular RH problem (\ref{regRH}). Accounting now
 for the relation
 \[
 \partial_t\det\psi_+(k)=-i\epsilon\,\mathrm{tr}H_+\det\psi_+(k),
 \]
 we eventually obtain a simple evolution equation for the zero
 $k_1$:
 \begin{eqnarray}\label{kk}
 \partial_tk_1&=&\frac{i}{2}\,\epsilon\,\mathrm{tr}\,\mathrm{Res}\left[H_+(k),k_1\right]\\
 &=&\frac{i}{2}\,
 \epsilon\,\mathrm{Res}\left[\Upsilon_{+11}(k)+\Upsilon_{+22}(k),k_1\right].\nonumber
 \end{eqnarray}

 Summarizing, Eqs. (\ref{RH-cont}), (\ref{10-1})--(\ref{10-4}), and
 (\ref{kk})
  determine perturbation-induced
 evolution of the RH data. It should be stressed that these equations
 are \textit{exact} because we did not yet refer to smallness of
 $\epsilon$ anywhere.
 At the same time, these equations cannot
 be directly applied because $\Upsilon_\pm$ entering them depend
 on unknown solutions $\psi_\pm$ of the spectral problem with the
 perturbed potential $\hat Q$. To proceed further, we develop,
 owing to the smallness of $\epsilon$, the
 adiabatic approximation of the general perturbation theory.

 \section{Adiabatic approximation}
 \label{sec:adia}

 In the framework of the adiabatic approximation, we assume
 that the perturbed soliton adjusts its shape
 to the unperturbed one at the cost of slow evolution of its
 parameters. Hence, only the discrete RH data are relevant in this
 approximation, and we can put $\tilde\psi_+=\openone$ for the
 solution of the regular RH problem (\ref{regRH}). Therefore,
 $\psi_+=\Xi$. In other words, it is the rational function
 $\Xi$ that completely determines soliton dynamics in the
 adiabatic approximation. In particular, we have
 \be\label{ups}
 \Upsilon_+=\int_{-\infty}^\infty\ud xE^{-1}\Xi^{-1}\hat
 R\,\Xi E, \quad \Xi^{-1}(k)=\Xi^\dag(k).
 \ee

 As an important example, we consider a perturbation caused by a
 small disturbance of the integrability condition (\ref{const}).
 In this case we introduce a small parameter as
 $\epsilon=c_0-c_2$, while the functional form of the
 perturbations $R_{\pm,0}$ has the form
 \be\label{Rform}
 R_{\pm,0}=\left(|\phi_+|^2+2|\phi_0|^2+|\phi_-|^2\right)\phi_{\pm,0}.
 \ee
 Inserting the explicit expressions for the
 soliton components
 $\phi_{\pm,0}$ (\ref{solr11}) into this equation gives
 \begin{eqnarray}
 R_+&=&(2\nu)^3e^{i(\varphi-\chi)}\cos^2\theta\,\mathrm{sech}^3z,\nonumber\\
 R_-&=&(2\nu)^3e^{i(\varphi+\chi)}\sin^2\theta\,\mathrm{sech}^3z,\label{pert}\\
 R_0&=&(2\nu)^3e^{i\varphi}\cos\theta\sin\theta\,\mathrm{sech}^3z.\nonumber
 \end{eqnarray}

 Matrix elements of $\Upsilon_+$ which are the main ingredients of
 the evolution equations for the soliton parameters are found from
 Eqs. (\ref{ups}) and (\ref{xi}),
 and the projector $P^{(1)}$ is calculated by means of the
 simple formula
 \[
 P^{(1)}=\frac{|1\rangle\langle 1|}{\langle 1|1\rangle}, \qquad
 \langle1|=|1\rangle^\dag,
 \]
 which follows from Eq. (\ref{proj}). The eigenvector $|1\rangle$ is
 given by Eq. (\ref{1}). As a result, matrix elements of the
 projector are as follows ($P_{ba}=P_{ab}^*$):
 \begin{eqnarray*}
 P_{11}^{(1)}\!\!&=&\!\!\frac{1}{2}e^z\cos^2\theta\,\mathrm{sech}z, \;
 P_{12}^{(1)}=\frac{1}{2}e^{z-i\chi}\cos\theta\sin\theta\,\mathrm{sech}z,\\
 P_{13}^{(1)}\!\!&=&\!\!\frac{1}{2}e^{i(\varphi-\chi)}\cos^2\theta\,\mathrm{sech}z,
 \;P_{24}^{(1)}=\frac{1}{2}e^{i(\varphi+\chi)}\sin^2\theta\,\mathrm{sech}z,\\
 P_{14}^{(1)}\!\!&=&\!\!P_{23}^{(1)}=\frac{1}{2}e^{i\varphi}\cos\theta\sin\theta\,\mathrm{sech}z,
 \;P_{22}^{(1)}=\frac{1}{2}\sin^2\theta\,\mathrm{sech}z,\\
 P_{33}^{(1)}\!\!&=&\!\!\frac{1}{2}e^{-z}\cos^2\theta\,\mathrm{sech}z,
 \; P_{44}^{(1)}=\frac{1}{2}e^{-z}\sin^2\theta\,\mathrm{sech}z,\\
 P_{34}^{(1)}\!&=&\!\frac{1}{2}e^{-z+i\chi}\cos\theta\sin\theta\,\mathrm{sech}z.
 \end{eqnarray*}

 Now we easily obtain from Eqs. (\ref{ups}) and (\ref{Rform}) that
 \[
 \mathrm{Res}\left[\Upsilon_{+11}(k)+\Upsilon_{+22}(k),k_1\right]=0.
 \]
 Therefore, $\partial_tk_1=0$ in accordance with Eq. (\ref{kk}), which means
 that the soliton amplitude and velocity preserve their initial
 values. Finding evolution of the other soliton parameters
 demands knowledge
 of the regular part of $\Upsilon_+(k_1)$. Calculation due to Eq.
 (\ref{regul}) gives
 \begin{eqnarray*}
 X_{11}&=&X_{12}=X_{21}=X_{22}=0,\\
 X_{31}&=&(2\nu^2)\exp(-\rho-i\varphi_\alpha+i\chi)\cos^2\theta,\\
 X_{42}&=&(2\nu^2)\exp(-\rho-i\varphi_\alpha-i\chi)\sin^2\theta,\\
 X_{41}&=&X_{32}=(2\nu^2)\exp(-\rho-i\varphi_\alpha)\cos\theta\sin\theta.
 \end{eqnarray*}
 Substituting these functions into Eqs. (\ref{10-1})--(\ref{10-4}),
 we obtain:
 \begin{eqnarray*}
 \rho&=&\mathrm{const}, \quad \theta=\mathrm{const}, \\
 \chi&=&\mathrm{const}, \quad
 \varphi_\alpha(t)=\varphi_\alpha(0)-4\epsilon\nu^2t.
 \end{eqnarray*}
 As a result, within the adiabatic approximation, the only
 manifestation of the perturbation caused by a small deviation from
 the integrability condition (\ref{const}) consists in a small
 shift of the soliton frequency equal to $4\epsilon\nu^2$. Hence, a
 ferromagnetic soliton is a pretty robust object against a small
 disturbance of the integrability condition.

\begin{figure}
   \begin{center}
   {\hspace{-4mm}\includegraphics[width=0.48\textwidth]{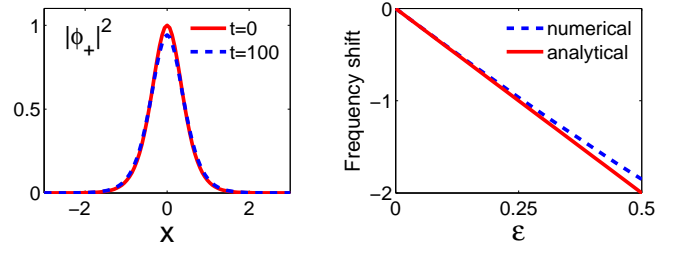}}
   \caption{(Color online) Left-hand panel: evolution of the perturbed $\phi_+$
   component profile obtained
   numerically. Here
   $\epsilon=0.1$. Right-hand
   panel: comparison of the analytically predicted frequency shift
   of the perturbed soliton with that obtained numerically from Eqs. (\ref{per-sys})
   and (\ref{Rform}) with $\epsilon=0.1$. In
   both panels $|\alpha|=|\beta|=|\gamma|=0.5$.
   }\label{fig3}
   \end{center}
   \end{figure}

 This conclusion has been checked by comparison with direct
 simulations of the perturbed equations (\ref{per-sys}). The
 left-hand
 panel of Fig. \ref{fig3} demonstrates the evolution of the
 perturbed
 $\phi_+$ component profile. We see a small profile distortion.
 Very little energy radiation is emitted to the far field. The same
 results are valid for the other two components. It is seen from the
 right-hand panel that there is a good agreement of the predicted linear
 dependence of the frequency shift on $\epsilon$ with that
 obtained numerically.

 \section{Conclusion}
 \label{sec:conclusion}

 In the present paper we have developed a perturbation theory for
 bright solitons of the integrable spinor BEC model.
 This model is equivalent to the $2\times2$ matrix NLS equation and
 is naturally associated with the matrix RH problem. We have
 demonstrated the efficiency of the formalism based on the RH
 problem, for solving both integrable and nearly integrable versions
 of the spinor BEC model. We have obtained the rank-one and rank-two
 soliton solutions of the model. Depending on the spin properties,
 the rank-two soliton can be of the ferromagnetic type or of the
 polar type. We have proven that the ferromagnetic soliton is
 equivalent to the rank-one soliton. As regards the polar soliton,
 its profile is characterized by a two-humped structure in a wide
 region of the soliton parameters. We have observed from numerical
 experiments that the polar soliton is unstable under the action of
 a perturbation and splits into a pair of ferromagnetic
 solitons. Owing to this fact, the problem to construct a
 perturbation theory for the spinor BEC solitons has been reduced to
 that for the rank-one solitons.

 We have derived perturbation-induced evolution equations for the
 soliton parameters. In the adiabatic approximation of the
 perturbation theory these equations have been applied to a
 practically important case of a perturbation caused by a small deviation of the
 model parameters from those in the integrable case. We have shown a
 considerable stability of the ferromagnetic soliton in the presence
 of such a perturbation. Namely, the soliton preserves its
 amplitude, velocity and spin properties, a small frequency shift
 being the only manifestation of the perturbed environment. At the same
 time, the polar soliton solution of the
 integrable model has a restrictive area of applicability due to its instability and
 splitting under perturbations.

 Three more points deserve a special comment. First, instability of
 a perturbed polar soliton and its splitting into ferromagnetic ones
 have been observed numerically. Analytical study of this phenomenon
 demands a separate consideration and can be performed, for example,
 by a stability analysis as developed in Ref. \cite{J2}. Second, we have
 restricted ourselves to the study of the adiabatic approximation of
 the general perturbation theory. Our equations permit us to go
 beyond this approximation and take into account the soliton shape
 distortion effects. However, quantitative characteristics of the
 first-order effects are too small to be verified experimentally, at
 least at present. Examples of practical calculations in the
 first-order approximation can be found in Ref. \cite{DMR}.
 Third, the formalism developed here for the single
 perturbed soliton can be straightforwardly generalized to the case of $N$
 weakly interacting solitons arranged into a train-like
 configuration. Analysis of the soliton train dynamics by the
 soliton perturbation theory
 can be found in Refs. \cite{GDY,ZY} for optical solitons and in Ref.
 \cite{Salerno} for scalar bright BEC solitons.

 \section{Acknowledgments}

 Constructive propositions of P. Kevrekidis are greatly appreciated.
 E.D. thanks the Department of Mathematics and Statistics of the University
 of Vermont for the
 hospitality.

 \end{document}